\documentstyle[aps]{revtex}
\input{epsf}

\newcommand{\eff}{\mbox{\scriptsize eff}}
\newcommand{\fac}{\mbox{\scriptsize fac}}
\newcommand{\MS}{\overline{\mbox{\tiny MS}}}

\begin{document}

\draft

\title{Factorization and  Nonfactorization in $B$~Decays}
\author{F. M. Al-Shamali and A. N. Kamal}
\address{Theoretical Physics Institute and Department of Physics,
University of Alberta, Edmonton, Alberta, Canada T6G 2J1. \\
\vspace{1.5mm} {\em Report no.\ : Alberta Thy 05-99}}
\date{\today}

\maketitle
\begin{abstract}
Using NLL values for Wilson coefficients and including the contributions from the penguin diagrams, we estimate the amount of nonfactorization in two-body hadronic $B$ decays. Also, we investigate the model dependence of the nonfactorization parameters by performing the calculation using different models for the form factors. The results support the universality of nonfactorizable contributions  in both Cabibbo-favored and Cabibbo-suppressed $B$~decays.
\end{abstract}



\section{Introduction}

The decay rates (and, in few cases, polarization) of a large number of hadronic channels have been experimentally measured to sufficient accuracy. However, as yet there is no reliable theoretical method to derive the corresponding amplitudes starting from the basic principles of the Standard Model. This is due to our inability to quantify the participation of strong interactions in such processes. The powerful theoretical tools of perturbation theory which are used in purely electroweak interactions are not very useful in situations involving strong interactions. Nonperturbative techniques, such as lattice calculations and QCD sum rules, are still under development. However, the asymptotic freedom property of QCD allows us to separate the gluon contribution, in a given process, into that due to high energy (hard) gluons and that  due to low energy (soft) gluons. The former contribution is relatively easy to compute using perturbation techniques and renormalization group equations. In fact, an impressive amount of work in this regard has been done \cite{ref:Gilman-79,ref:Ponce-81,ref:Ruckl-83,ref:Buras-90,ref:Buras-92,ref:Buchalla-96}, where hard gluon effects were parametrized through Wilson coefficients which have been calculated up to next-to-leading logarithmic order (NLL). It is the soft gluon contribution which is difficult to handle and constitutes the main source of uncertainty in hadronic weak decays.

\begin{figure}
\epsfxsize=7.0in  \centerline{\epsffile{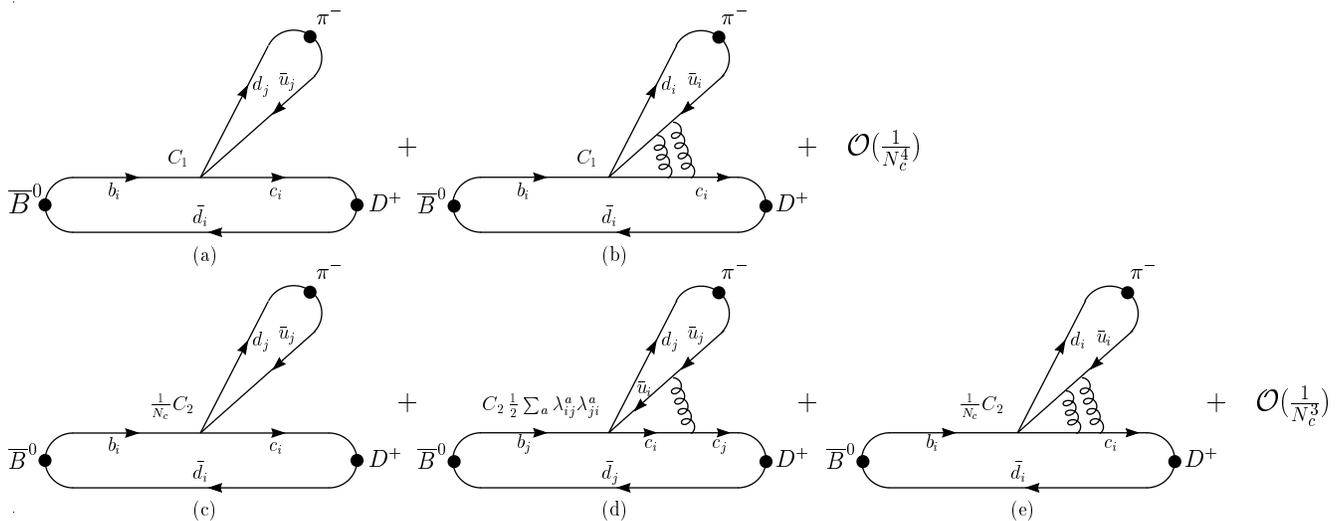}}

\caption{The first few terms of $1/N_c$~expansion, shown in a diagramatic language, that contribute to the decay amplitude of the process $\overline{B}^0 \rightarrow D^+ \pi^-$.}
\label{fig:diagrams-I}
\end{figure}

\begin{figure}
\epsfxsize=2.5in  \centerline{\epsffile{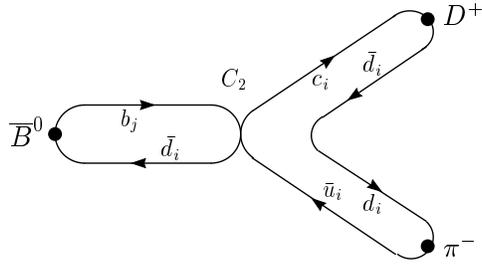}}

\caption{The annihilation diagram contributing to the decay amplitude of the process $\overline{B}^0 \rightarrow D^+ \pi^-$.}
\label{fig:ann-diagram}
\end{figure}

In general, the effective Hamiltonian in the hadronic decays of $B$ mesons takes the form $H_{\eff} \sim \sum_i C_i Q_i$ where $C$s are the Wilson coefficients that contain the hard gluon (or short distance) effects and $Q$s are four-quark operators that are products of two Dirac currents. In calculating the decay amplitudes for two-body hadronic decays we then encounter matrix elements of the form $\langle f_1 \, f_2 | Q | i \rangle$, where $i$ is the initial state particle, $f_1$ and $f_2$ are the final-state particles. For example, the effective Hamiltonian relevant to the process $\overline{B}^0 \rightarrow D^+ \pi^-$ is given by (see the next section for details),
\begin{equation}
{\cal H}_{\eff} =  \frac{G_F}{\sqrt{2}} V_{cb} V_{ud}^*
\left[ C_1 \, Q_1 + C_2 \, Q_2 \right] ,
\label{eq:Heff-bcud}
\end{equation}
where
\begin{eqnarray}
Q_1 & = & (\bar{c_i} \, b_i)_L \, (\bar{d_j} \, u_j)_L , \\
Q_2 & = & (\bar{c_i} \, b_j)_L \, (\bar{d_j} \, u_i)_L \nonumber\\
& = & \frac{1}{N_c} (\bar{c_i} \, b_i)_L \, (\bar{d_j} \, u_j)_L + \frac{1}{2} (\bar{c_i} \lambda_{ij}  b_j)_L \, (\bar{d_k} \lambda_{kl}  u_l)_L .
\label{eq:fierz}
\end{eqnarray}
The subscripts $i$, $j$, $k$, and $l$ represent color, $N_c$ is the number of colors, $\lambda$s are the Gell-Mann matrices and $L$ represents a left-handed current,
\begin{equation}
(\bar{c_i} \, b_i)_L = \bar{c}_i \gamma^\mu (1 - \gamma_5) b_i .
\end{equation}
In a diagrammatic language similar to that used in Ref.~\cite{ref:Buras-86}, we show in Fig.~\ref{fig:diagrams-I} different diagrams contributing to the decay amplitude of this process. Regarding these diagrams several remarks can be made. First, the leading diagram generated by $Q_1$ (Fig.~\ref{fig:diagrams-I}~(a)) is factorizable, where all soft gluon exchanges inside the closed loops are accounted for by the $B \rightarrow D$ form factors and the pion decay constant. Second, the next-to-leading diagram generated by $Q_1$ (Fig.~\ref{fig:diagrams-I}~(b)) is not factorizable and is of the order of $1/N_c^2$ relative to the leading one. This is due to the double soft-gluon exchange between two color-singlet mesons. Third, when the operator $Q_2$ is rewritten using Fierz transformation and color algebra in the form of Eq.~(\ref{eq:fierz}), it generates, the lower set of diagrams in Fig.~\ref{fig:diagrams-I}. The diagram in Fig.~\ref{fig:diagrams-I}~(c) which is generated by the color-singlet current is factorizable whereas those in Figs.~\ref{fig:diagrams-I}~(d) and \ref{fig:diagrams-I}~(e) are not. Finally, the annihilation diagram in Fig.~\ref{fig:ann-diagram} is generated by $Q_2$ alone.

As we said earlier the factorizable part can be parametrized in terms of form factors and decay constants. On the other hand, the nonfactorizable parts are more difficult to handle. A popular method is to omit the nonfactorizable parts altogether and to assume complete factorization of the decay amplitude. Some justification for this approach is provided by the fact that the nonfactorizable diagrams are nonleading in the $1/N_c$ expansion. However, there are two weak points in this argument. First, $1/N_c$ (where $N_c = 3$) is not a good expansion parameter. Second, even the factorizable part carries some uncertainties due to the model dependence of the form factors and scale dependence of the Wilson coefficients. Of course, experiment has the final say regarding the validity of such an approach.

In $B$~decays into two psedoscalar mesons, the annihilation diagram contribution can be argued to be small. For example, in the process $\overline{B}^0 \rightarrow D^+ \pi^-$ we expect the ratio of the annihilation diagram to the factorizable tree diagram to be of the order (here the annihilation diagram is also calculated in the factorization approximation)
\begin{equation}
\sim \; \frac{C_2}{C_1} \; \frac{(m_D^2 - m_\pi^2)}{(m_B^2 - m_D^2)} \; \frac{f_B}{f_\pi} \; \frac{F^{D \rightarrow \pi}(m_B^2)}{F^{B \rightarrow D}(m_\pi^2)} \approx \; 0.03 \; \frac{f_B}{f_\pi} \; \frac{F^{D \rightarrow \pi}(m_B^2)}{F^{B \rightarrow D}(m_\pi^2)},
\end{equation}
where the values of the Wilson coefficients used are shown in
Table~\ref{tab:Wilson-LL-NLL}. So, unless $F^{D \rightarrow \pi}(m_B^2)$ has a very high value (which is not expected) the contribution of the annihilation diagram is, in all likelihood, very small. In this paper, we neglect the annihilation terms in decays involving vector particles also. 

At this point we should differentiate between two approaches to factorization. In the first approach, called naive factorization (see for example Ref.~\cite{ref:Wirble-85}), both leading and subleading diagrams are included in amplitude calculations. In the second approach (see for example Ref.~\cite{ref:Buras-86}), called the large $N_c$ limit, only the leading diagrams in $1/N_c$ are included. If the leading diagram in a decay process is generated by the operator $Q_1$, factorization works at its best. This is because the subleading diagrams are suppressed further by the Wilson coefficient $C_2$. In the large $N_c$ limit, the decay amplitude for such processes, which are labelled class~I \cite{ref:Wirble-85}, is proportional to
\begin{equation}
C_1 \, \langle f_1 | J^\mu | i \rangle  \langle f_2 | J'_\mu | 0 \rangle
\label{eq:classI-Nlimit}
\end{equation}
In this equation $J^\mu$ and $J'_\mu$ are two Dirac currents. In naive factorization the amplitude is proportional to
\begin{equation}
\left(C_1 + \frac{C_2}{N_c} \right) \, \langle f_1 | J^\mu | i \rangle  \langle f_2 | J'_\mu | 0 \rangle .
\label{eq:classI-Nnaive}
\end{equation}
The difference between decay rates predicted by these two amplitudes (defined by Eqs.~(\ref{eq:classI-Nlimit}) and (\ref{eq:classI-Nnaive}))is not very large. It is about 20\% (15\%) in the case of $D$ ($B$)~decays. However, in class~II processes, where $Q_2$ generates the leading factorizable diagram, we find that the decay amplitudes are proportional to
\begin{equation}
C_2 \, \langle f_2 | J^\mu | i \rangle  \langle f_1 | J'_\mu | 0 \rangle
\label{eq:classII-Nlimit}
\end{equation}
and
\begin{equation}
\left(C_2 + \frac{C_1}{N_c} \right) \, \langle f_2 | J^\mu | i \rangle  \langle f_1 | J'_\mu | 0 \rangle
\label{eq:classII-Nnaive}
\end{equation}
in the large $N_c$ limit and in the naive factorization, respectively. Unlike processes of class~I, the inclusion of the subleading factorizable diagram in class~II processes has a significant effect on the decay amplitude. For example, in the case of $D$~decays the predicted decay rates drop by one order of magnitude. In class~III processes, both $Q_1$ and $Q_2$ generate leading and subleading factorizable diagrams. So, the amplitude is proportional to
\begin{equation}
C_1 \, \langle f_1 | J^\mu | i \rangle  \langle f_2 | J'_\mu | 0 \rangle + C_2 \, \langle f_2 | J^\mu | i \rangle  \langle f_1 | J'_\mu | 0 \rangle
\label{eq:classIII-Nlimit}
\end{equation}
and
\begin{equation}
\left(C_1 + \frac{C_2}{N_c} \right) \, \langle f_1 | J^\mu | i \rangle  \langle f_2 | J'_\mu | 0 \rangle + \left(C_2 + \frac{C_1}{N_c} \right) \, \langle f_2 | J^\mu | i \rangle  \langle f_1 | J'_\mu | 0 \rangle
\label{eq:classIII-Nnaive}
\end{equation}
in the large $N_c$ limit and in the naive factorization, respectively.

When factorization was applied in studying the decays of $D$~mesons, it was found that the large $N_c$ limit worked much better than naive factorization. This seemed logical from the point of view of $1/N_c$ expansion because \cite{ref:Buras-86} it is not justified to include only part of the terms in a certain order. This led to the anticipation that $1/N_c$ expansion can also be used to describe $B$~decays. However, when $B$~measurements started to appear it was found that both factorization approaches failed to describe the new experimental results, especially for class~II and class~III processes. For example, the decay rate of the process $B \rightarrow K J/\psi$ was found to be about one order of magnitude higher than that predicted by factorization \cite{ref:Gourdin-94,ref:Aleksan-95}. This shifted the attention of many physicists towards the nonfactorizable terms, which were previously omitted. The argument used is that these terms are not negligible and they contribute significantly to the decay amplitude. However, in the case of $D$~decays it just happened that the subleading factorizable diagram almost canceled out the nonfactorizable one due to a destructive interference between the two. On the other hand, in the case of $B$~decays the interference seemed to be constructive.

Ideally, if the nonfactorizable contributions could be calculated from the basic principles of the Standard Model we could systematically test the validity of the $1/N_c$ expansion. Unfortunately, with the present theoretical capabilities this is a difficult task (see Refs.~\cite{ref:Blok-93,ref:Ruckl-98} for some attempts in this direction). However, until such computational methods become more reliable, we can still do phenomenological parameterizations that can be compared with experimental data.

So, in going one step beyond factorization we can write the decay amplitude up to the first order in $1/N_c$ but including the nonfactorizable diagram generated by color-octet currents. Changes would then be made to the amplitudes in (\ref{eq:classI-Nnaive}), (\ref{eq:classII-Nnaive}) and (\ref{eq:classIII-Nnaive}) through the replacements:
\begin{eqnarray}
\left(C_1 + \frac{C_2}{N_c} \right) & \rightarrow &
\left(C_1 + \frac{C_2}{N_c} + C_2 \, \varepsilon_8 \right) , \nonumber\\
\left(C_2 + \frac{C_1}{N_c} \right) & \rightarrow &
\left(C_2 + \frac{C_1}{N_c} + C_1 \, \varepsilon_8 \right) ,
\label{eq:a1eff-a2eff}
\end{eqnarray}
where $\varepsilon_8$ represents the contribution of the color octet nonfactorizable diagram relative to the factorizable one. In Refs.~\cite{ref:Cheng-94,ref:Kamal-94,ref:Soares-95,ref:Neubert-97} several authors attempted to estimate the amount of nonfactorizable contributions in several decay channels of $D$ and $B$~decays. The results highlight  two main points. First, in $D$ decays the value of $\varepsilon_8$ is negative and highly process dependent. Second, in $B$ decays $\varepsilon_8$ is positive and is not as process dependent as in $D$ decays. In addition, the results obtained in Ref.~\cite{ref:Cheng-94} seems to satisfy the inequality
\begin{equation}
\left| \varepsilon_8(D \rightarrow V P) \right| >
\left| \varepsilon_8(D \rightarrow P P) \right| >
\left| \varepsilon_8(B \mbox{ decays}) \right| ,
\end{equation}
which finds some justification in the color transparency argument \cite{ref:Bjorken-89}. According to this argument soft-gluon exchange becomes less important as the final products of an interaction become more energetic. This is because the final state particles leave the interaction region very quickly allowing little time for final state interactions. It should be mentioned here that some authors\cite{ref:Ali-98-1,ref:Ali-98-2,ref:Cheng-98} treated $N_c$ as a free parameter whose effective value is used to indicate the amount of nonfactorization in a decay process.

One of the issues that arises when parameterizing nonfactorizable contributions is how to handle processes with two vector mesons in the final state? This is because it is not clear whether or not the three Lorentz scalar structures of the decay amplitude should receive the same contribution from the nonfactorizable terms, i.e.\ does nonfactorization lead verily to an over all factor? This issue was discussed in Refs.~\cite{ref:Kamal-94,ref:Cheng-94}. In Ref.~\cite{ref:Shamali-98} we tackled this issue in some detail for the process $B \rightarrow J/\psi K^*$. Using a full amplitude measurement \cite{ref:CLEO-97} by CLEO, the amount  of nonfactorizable contribution to each of the three Lorentz-scalar structures was calculated in five different models for the form factors. The results allowed an explanation of the experimental data using equal amount of nonfactorization in each part of the Lorentz amplitude, implying that an overall nonfactorization factor was adequate.

Assuming universality (process-independence) of the nonfactorization parameters in $B$~decays, we estimated their values in Ref.~\cite{ref:Shamali-99} using a more definitive calculation. However, for the Wilson coefficients we used the values calculated up to leading logarithmic order (LL) and neglected all contributions from the penguin diagrams. Also, all the calculations were done using only one model for the form factors (BSW~II model). The results supported the proposition of universality of the nonfactorization parameters in Cabibbo-favored $B$~decays. However, a number of questions were also raised. First, how will the results change if we use the Wilson coefficients calculated up to NLL? Second, how do we parameterize nonfactorization generated by penguin diagrams and how important they are? Third, how much model-dependence is there in the estimated nonfactorization parameters? Finally, can we extend the proposed universality of the nonfactorization parameters to include Cabibbo-suppressed $B$~decays? These are the questions we try to address in this paper.

The paper is arranged as follows in Sections~II and III we present the Wilson coefficients and CKM matrix elements used in the calculations. In Section~IV we calculate the effects of penguin diagrams and NLL Wilson coefficients on the predictions of naive factorization. In Section~V the nonfactorization parameters are estimated in five models for the form factors. In Section~VI we show branching ratio predictions of several sets of Cabibbo-favored and Cabibbo-suppressed $B$~decays. The last Section is a discussion of the results and a conclusion.


\section{Wilson Coefficients in NLL}

\begin{figure}
\epsfxsize=2.5in  \centerline{\epsffile{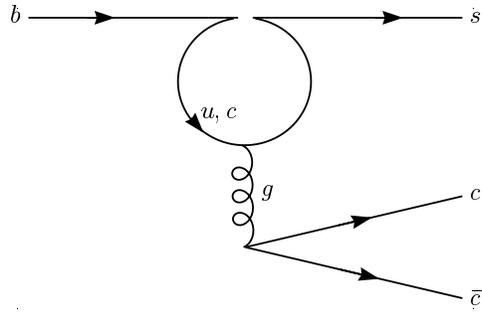}}

\caption{The Penguin Feynman diagrams for the process $b \rightarrow c \bar{c} s$ in the effective theory.}
\label{fig:feyn}
\end{figure}

In the absence of strong interactions, the effective Hamiltonian for the process $b \rightarrow c \bar{c} s$, is given by
\begin{equation}
{\cal H}_{\eff} =  \frac{G_F}{\sqrt{2}} V_{cb} V_{cs}^*
( \bar{c}_i b_i )_L \, ( \bar{s}_j c_j )_L .
\label{eq:Heff-bccs}
\end{equation}
When QCD effects are included, the contribution of the penguin diagram in Fig.~\ref{fig:feyn} should be considered beside the current $\times$ current diagrams. As a result, the effective Hamiltonian generalizes to \cite{ref:Ali-98-1,ref:Cheng-98}
\begin{eqnarray}
{\cal H}_{\eff} & = & \frac{G_F}{\sqrt{2}} \left[ \rule{0mm}{6mm}
V_{ub} V_{us}^* \left( C_1 \, Q^u_1 + C_2 \, Q^u_2\right)
+ V_{cb} V_{cs}^* \left( C_1 \, Q^c_1 + C_2 \, Q^c_2\right) + \left( V_{ub} V_{us}^* + V_{cb} V_{cs}^* \right)
\sum^6_{i = 3} C_i \, Q_i \right] ,
\label{eq:Heff-bccs-QCD}
\end{eqnarray}
where
\begin{eqnarray}
Q^u_1 & = & (\bar{u}_i b_i)_L \, (\bar{s}_j u_j)_L \nonumber\\
Q^u_2 & = & (\bar{u}_i b_j)_L \, (\bar{s}_j u_i)_L \nonumber\\
Q^c_1 & = & (\bar{c}_i b_i)_L \, (\bar{s}_j c_j)_L \nonumber\\
Q^c_2 & = & (\bar{c}_i b_j)_L \, (\bar{s}_j c_i)_L \nonumber\\ 
Q_3 & = & (\bar{s}_i b_i)_L \, \sum_q (\bar{q}_j q_j)_L \label{eq:Qs-bccs}\\
Q_4 & = & (\bar{s}_i b_j)_L \, \sum_q (\bar{q}_j q_i)_L \nonumber\\
Q_5 & = & (\bar{s}_i b_i)_L \, \sum_q (\bar{q}_j q_j)_R \nonumber\\
Q_6 & = & (\bar{s}_i b_j)_L \, \sum_q (\bar{q}_j q_i)_R \,. \nonumber
\end{eqnarray}
The superscript $R$ represents a right-handed current such that
\begin{equation}
(\bar{q_j} \, q_j)_R = \bar{q}_j \gamma^\mu (1 + \gamma_5) q_j .
\end{equation}
Even though, the local operators $Q^u_1$ and $Q^u_2$ don't contribute to processes of the type $b \rightarrow c \bar{c} s$ through tree diagrams, they do contribute through penguin diagrams. In writing (\ref{eq:Heff-bccs-QCD}), we have made use of the following unitarity condition of the CKM matrix elements
\begin{equation}
V_{ub} V_{us}^* + V_{cb} V_{cs}^* + V_{tb} V_{ts}^* = 0 .
\end{equation}

The Wilson Coefficients, $C_1$ and $C_2$, are determined by requiring the effective Hamiltonian to reproduce the amplitude calculated in the full theory. In LL calculations this matching of the full theory onto the effective theory is done by keeping only the terms containing logarithmic corrections of order $\alpha_s \ln( )$. In NLL calculations, the terms containing constant contributions of  order $\alpha_s$ are kept in addition to the terms of order $\alpha_s \ln( )$. The Wilson coefficients turn out to be regularization scheme dependent. However, if the matrix elements $\left< Q_i (\mu) \right>$ are evaluated at the same scheme as the Wilson coefficients $C_i (\mu)$, the scheme dependence cancels out. From Eq.~(\ref{eq:Heff-bccs-QCD}), we see that the decay amplitude in the effective theory has the form
\begin{equation}
{\cal A}_{\eff} \propto C_i (\mu) \; \left< Q_i (\mu) \right>.
\label{eq:Aeff-Cmu-Qmu}
\end{equation}
If these matrix elements are related to the tree matrix elements by
\begin{equation}
\left< Q_i (\mu) \right> = g_{ij}(\mu) \, \left< Q_j\right>^{\mbox{\scriptsize tree}}  ,
\label{eq:Qmu-g-Qtree}
\end{equation}
then we can write the effective amplitude as 
\begin{eqnarray}
{\cal A}_{\eff} & \propto &
C_i(\mu) \; g_{ij}(\mu) \left< Q_j\right>^{\mbox{\scriptsize tree}} , \nonumber\\
& \propto & C_i^{\eff} \; \left< Q_i\right>^{\mbox{\scriptsize tree}} .
\label{eq:Aeff-Ceff-Qtree}
\end{eqnarray}
At the quark level, the scale and scheme dependence of $\left< Q_i (\mu)\right>$, which is carried by $g(\mu)$, cancel the scale and scheme dependence of the Wilson coefficients \cite{ref:Ali-98-1,ref:Cheng-98,ref:Fleischer-93,ref:Kramer-94}. So, both $C_i^{\eff}$ and $\left< Q_i\right>^{\mbox{\scriptsize tree}}$ are scale and scheme independent.

From above, we can write the decay amplitude as
\begin{eqnarray}
\langle {\cal H}_{\eff} \rangle = \frac{G_F}{\sqrt{2}} \left[ \rule{0mm}{6mm} V_{cb} V_{cs}^* \left( C^{\eff}_1 \langle Q^c_1 \rangle^{\mbox{\scriptsize tree}} + C^{\eff}_2 \langle Q^c_2 \rangle^{\mbox{\scriptsize tree}} \right) + \sum^6_{i = 3} \left(V_{ub} V_{us}^* \, C^{u \,\eff}_i + V_{cb} V_{cs}^* \, C^{c \,\eff}_i \right)
\langle Q_i \rangle^{\mbox{\scriptsize tree}} \right],
\label{eq:Aeff-bccs}
\end{eqnarray}
where the penguin contributions from $\langle Q^u_{i = 1, 2} \rangle$ and $\langle Q^c_{i = 1, 2} \rangle$ are included in $C^{u \,\eff}_{i = 3, \ldots, 6}$ and $C^{c \,\eff}_{i = 3, \ldots, 6}$ , respectively. Using the conventions of Ref.~\cite{ref:Ali-98-1}, the explicit forms of $C_i^{\eff}$ are given by
\begin{eqnarray}
C_1^{\eff} & = & C_1 +
\frac{\alpha_s}{4 \pi} \left( r_V^T + \gamma_V^T \ln \frac{m_b}{\mu} \right)_{1j} C_j , \nonumber\\
C_2^{\eff} & = & C_2 +
\frac{\alpha_s}{4 \pi} \left( r_V^T + \gamma_V^T \ln \frac{m_b}{\mu} \right)_{2j} C_j , \nonumber\\
C_3^{q \,\eff} & = & C_3 - \frac{\alpha_s}{24 \pi} \left[ C_t(m_q) + C_p \right] + \frac{\alpha_s}{4 \pi} \left( r_V^T + \gamma_V^T \ln \frac{m_b}{\mu} \right)_{3j} C_j \nonumber\\
C_4^{q \,\eff} & = & C_4 + \frac{\alpha_s}{8 \pi} \left[ C_t(m_q) + C_p \right] + \frac{\alpha_s}{4 \pi} \left( r_V^T + \gamma_V^T \ln \frac{m_b}{\mu} \right)_{4j} C_j \label{eq:C1-C6-NLL-eff} \\
C_5^{q \,\eff} & = & C_3 - \frac{\alpha_s}{24 \pi} \left[ C_t(m_q) + C_p \right] \frac{\alpha_s}{4 \pi} \left( r_V^T + \gamma_V^T \ln \frac{m_b}{\mu} \right)_{5j} C_j \nonumber\\
C_6^{q \,\eff} & = & C_4 + \frac{\alpha_s}{8 \pi} \left[ C_t(m_q) + C_p \right] \frac{\alpha_s}{4 \pi} \left( r_V^T + \gamma_V^T \ln \frac{m_b}{\mu} \right)_{6j} C_j \;, \nonumber
\end{eqnarray}
where the superscript $q$ represent either the $u$ or the $c$ quark.  In (\ref{eq:C1-C6-NLL-eff}), the quantities $C_t(m_q)$ and $C_p$ arise from the penguin-like diagrams of  the operators $Q^q_{i = 1, 2}$ and $Q_{i = 3, \ldots,6}$, respectively, whereas the matrix $\left( r_V + \gamma_V \ln \frac{m_b}{\mu} \right)$ is due to the vertex and self-energy corrections to the operators $Q_{i = 1, \ldots,6}$. These quantities are given by \cite{ref:Ali-98-1}
\begin{equation}
r_V = \left (\matrix{
{7 \over 3} & -7 & 0 & 0 & 0 & 0 \cr
-7 & {7 \over 3} & 0 & 0 & 0 & 0 \cr
0 & 0 & {7 \over 3} & -7 & 0 & 0 \cr
0 & 0 & -7 & {7 \over 3} & 0 & 0 \cr
0 & 0 & 0 & 0 & -{1 \over 3} & 1 \cr
0 & 0 & 0 & 0 & -3 & {35 \over 3} \cr }\right ) ,
\end{equation}
\begin{equation}
\gamma_V = \left (\matrix{
-2 & 6 & 0 & 0 & 0 & 0 \cr
6 & -2 & 0 & 0 & 0 & 0 \cr
0 & 0 & -2 & 6 & 0 & 0 \cr
0 & 0 & 6 & -2 & 0 & 0 \cr
0 & 0 & 0 & 0 & 2 & -6 \cr
0 & 0 & 0 & 0 & 0 & -16 \cr }\right ) ,
\end{equation}
\begin{eqnarray}
C_t(m_q) & = & C_1(\mu) \left[ {2 \over 3} + {2 \over 3} \ln \frac{m^2_q}{\mu^2} - \Delta F_1 \left(\frac{k^2}{m^2_q}\right) \right] ,\\
C_p & = & C_3(\mu) \left[ {4 \over 3} + {2 \over 3} \ln \frac{m_s^2}{\mu^2}  +
{2 \over 3} \ln \frac{m_b^2}{\mu^2}  - \Delta F_1 \left(\frac{k^2}{m_s^2}\right) \Delta F_1 \left(\frac{k^2}{m_b^2}\right)\right] \nonumber\\
&& + \left[ C_4(\mu) + C_6(\mu) \right] \sum_{i = u, d, s, c, b} \left[ {2 \over 3} \ln \frac{m_i^2}{\mu^2} - \Delta F_1 \left(\frac{k^2}{m_i^2}\right) \right] ,
\end{eqnarray}
and
\begin{equation}
\Delta F_1(z) = - 4 \int_0^1 dx \; x (1 - x) \ln [ 1 - z \, x (1 - x)] .
\end{equation}
Here, $k$ is the momentum carried by the gluon in Fig.~\ref{fig:feyn}\@. 

The calculation of the penguin-driven amplitudes in the factorization assumption involves additional assumptions, an effective value of $k^2$ for example. In a complete calculation \cite{ref:Kamal-97} $k^2$ would not be a variable; it would be integrated over the wave functions of the hadrons with its own uncertainties. In the absence of a complete knowledge of the hadronic wave functions, the choice is either to select $k^2$ judiciously or to admit new unknowns through the hadronic wave functions. In penguin calculations, one generally opts for the first alternative and chooses $k^2$ in the range: $m_b^2/4 \leq k^2 \leq m_b^2/2$. In the calculations presented here we have chosen $k^2 = m_b^2/2$.

Working in the naive dimensional regularization (NDR) scheme, the Wilson coefficients at the scale $\mu =$ 4.6~GeV are listed in Table~\ref{tab:Wilson-LL-NLL}. Using the definitions in Eq.~(\ref{eq:C1-C6-NLL-eff}) and the values of $C_i(\mu)$ in Table~\ref{tab:Wilson-LL-NLL} we calculate in Table~\ref{tab:Ceff-NLL} the values of the effective Wilson coefficients $C_i^{\eff}$. For $C_3^{\eff} - C_6^{\eff}$, which include the contributions from the QCD penguin diagrams, we list two sets of values, depending on which quark is in the loop (see Fig.~\ref{fig:feyn}). For quark masses, we used the the following running values at the $b$-quark mass scale \cite{ref:Fusaoka-98}: $m_u = 3.17$ MeV, $m_d = 6.37$ MeV,  $m_s = 0.127$ GeV, $m_c = 0.949$ GeV, $m_b = 4.34$ GeV and $m_t = 170$~GeV\@.


\section{CKM Matrix Elements}

To be able to calculate the decay amplitudes of the processes considered in this work, we need the values of the Cabibbo-Kobayashi-Maskawa (CKM) matrix elements. In Wolfenstein parametrization \cite{ref:Wolfenstein-83}, and up to the fourth order in $\lambda$, these elements take the form
\begin{equation}
\left( \begin{array}{ccc}
V_{ud} & V_{us} & V_{ub} \\
V_{cd} & V_{cs} & V_{cb} \\
V_{td} & V_{ts} & V_{tb} \end{array} \right) =
\left( \begin{array}{ccc}
1 - \frac{1}{2} \lambda^2 & \lambda & A \lambda^3 (\rho - i \eta) \\
-\lambda (1 + i A^2 \lambda^4 \eta) & 1 - \frac{1}{2} \lambda^2 & A \lambda^2 \\
 A \lambda^3 (1 - \rho - i \eta) & - A \lambda^2 & 1 \end{array} \right) .
\label{eq:Vckm-Wolfenstein}
\end{equation}
In a recent update \cite{ref:Ali-97}, the four parameters in Eq.~(\ref{eq:Vckm-Wolfenstein}) were given the values $\lambda = 0.2205$, $A = 0.81$, $\rho = 0.05$, and $\eta = 0.36$ using a $\chi^2$ fit to experimental data. The needed CKM matrix elements would then be:
\begin{eqnarray}
&& V_{ud} = 0.976 ,\; V_{us} = 0.221 ,\; V_{ub} = 0.00316 \; e^{-1.43 i} , \nonumber\\
&& V_{cd} =  -0.221 ,\; V_{cs} = 0.976 ,\; V_{cb} = 0.0394 .
\end{eqnarray}


\section{Decay Rates in Naive Factorization}

\subsection{Type $b \rightarrow c \bar{u} d$ Decays}

Decays of type $b \rightarrow c \bar{u} d$ proceed via the effective Hamiltonian presented in Eq.~(\ref{eq:Heff-bcud}), where only tree diagrams contribute to the decay amplitudes. In processes of class~I (for example $\overline{B}^0 \rightarrow D^+ \pi^-$) the factorizable part of the amplitude is proportional to $a_1 = C_1 + C_2/3$. As can bee seen from Tables~\ref{tab:Wilson-LL-NLL} and \ref{tab:Ceff-NLL}, the value of this parameter in LL ($a^{\mbox{\scriptsize LL}}_1 = 1.032$) is almost the same as that in NLL ($a^{\mbox{\scriptsize NLL}}_1 = 1.036$). Since the decay rates are proportional to $\left|a_1\right|^2$, the above values for $a_1$ gives a difference ($\Delta {\cal B}_{\fac}$) of less than $1\%$ between the branching ratios calculated using the Wilson coefficients in LL and the branching ratios calculated using the Wilson coefficients in NLL (see Table~\ref{tab:NLL-Ping-fac}). In processes of class~II (for example $\overline{B}^0 \rightarrow D^0 \pi^0$), the factorizable part of the decay amplitude is proportional to $a_2 = C_1/3 + C_2$. This parameter takes the values ($a^{\mbox{\scriptsize LL}}_2 =  0.090$) and ($a^{\mbox{\scriptsize NLL}}_2 = 0.059$) in LL and NLL, respectively. As a result, the predicted branching ratios (see Table~\ref{tab:NLL-Ping-fac}) in the naive factorization approximation, drop by about $57\%$ when working in NLL\@.

The decay amplitudes for class~III processes (for example $B^- \rightarrow D^0 \pi^-$) receive contributions from two tree diagrams causing a dependence on both $a_1$ and $a_2$. Therefore, $\Delta {\cal B}_{\fac}$ varies from one process to another in this class.  However, as can be seen from Table~\ref{tab:NLL-Ping-fac} these changes are relatively small (less that $6\%$). This is caused by the dominance of the part of the amplitude proportional to $a_1$ over that proportional to $a_2$. It should be mentioned here that, unlike the other two classes, $\Delta {\cal B}_{\fac}$ in class~III processes is model dependent and the values presented in Table~\ref{tab:NLL-Ping-fac} were calculated based on the BSW~II model, to be introduced later in this paper. However, due to the dominance of one part of the decay amplitude over the other, this model dependence is not very strong. 


\subsection{Type $b \rightarrow c \bar{c} s$ Decays}

Decays of type $b \rightarrow c \bar{c} s$, receive contributions from both tree and penguin diagrams and the relevant effective Hamiltonian is given by Eq.~(\ref{eq:Heff-bccs-QCD}). Consider the following four class~I decay channels, $\overline{B}^0 \rightarrow D^+ D_s^-$, $D^+ D_s^{*-}$, $D^{*+} D_s^-$ and $D^{*+} D_s^{*-}$. In order to extract the factorizable contributions to these processes, an appropriate transformation is needed for the operators $Q_5$ and $Q_6$ containing right-handed currents. Starting by $Q_5$, we can use Fierz transformation and color algebra to rewrite it as \cite{ref:Ali-98-1}
\begin{eqnarray}
Q_5 & = & (\bar{s} b)_L \, (\bar{c} c)_R \nonumber\\ & = & \left[
\bar{s}_i \gamma_\mu (1 - \gamma_5) b_i \right] \, \left[ \bar{c}_j
\gamma^\mu (1+ \gamma_5)c_j \right] \nonumber\\ & = & -2 \left[
\bar{c}_j (1 - \gamma_5) b_i \right] \, \left[ \bar{s}_i (1 +
\gamma_5) c_j \right] \nonumber\\ & = & -\frac{2}{3} \left[ \bar{c} (1
- \gamma_5) b \right] \, \left[ \bar{s} (1 + \gamma_5) c \right]
- \left[ \bar{c} \lambda^a (1 - \gamma_5) b \right] \,
\left[ \bar{s} \lambda^a (1 + \gamma_5) c \right] . \label{eq:Q5-Fierz}
\end{eqnarray}
For $Q_6$, we just use Fierz transformation to write it in the form
\begin{eqnarray}
Q_6 & = & (\bar{s}_i b_j)_L \, (\bar{c}_j c_i)_R \nonumber\\ & = &
\left[ \bar{s}_i \gamma_\mu (1 - \gamma_5) b_j \right] \, \left[
\bar{c}_j \gamma^\mu (1+ \gamma_5)c_i \right] \nonumber\\ & = & -2
\left[ \bar{c} (1 - \gamma_5) b \right] \, \left[ \bar{s} (1 +
\gamma_5) c \right] .
\end{eqnarray}
However, from Dirac Equation we can easily derive the following
relations between $(S + P)$ and $(S - P)$ Dirac bilinears, and $V$ and
$A$ bilinears
\begin{eqnarray}
\left[ \overline{s} (1 + \gamma_5) c \right] & = & i \left[
\frac{\partial^\mu (\overline{s} \gamma_\mu c)}{m_c - m_s} -
\frac{\partial^\mu (\overline{s} \gamma_\mu
\gamma_5 c)}{m_c+ m_s} \right], \\
\left[ \overline{c} (1 - \gamma_5) b \right] & = & i \left[ \frac{\partial^\mu (\overline{c} \gamma_\mu b )}{m_b - m_c} + \frac{\partial^\mu \left( \overline{c} \gamma_\mu \gamma_5 b \right)}{m_b+ m_c} \right] . \label{eq:cbL-current}
\end{eqnarray}
Using the relations (\ref{eq:Q5-Fierz}-\ref{eq:cbL-current}), we can write (assuming naive factorization, symbolized by the subscript~fac) the decay amplitude for the processes mentioned above as
\begin{eqnarray}
{\cal A}_{\fac} (\overline{B}^0 \rightarrow D^+ D_s^-) & = &
\frac{G_F}{\sqrt{2}} \left[ \rule{0mm}{6mm} V_{cb} V_{cs}^* \, a_1 + \sum_{q = u, c} V_{qb} V_{qs}^* \left(a_4^q + 2 a^q_6
\frac{m_{D_s}^2}{(m_b - m_c) (m_c + m_s)} \right) \right] \nonumber\\
&& \times \; \langle D^+ | (\overline{c} b)_L | \overline{B}^0
\rangle \, \langle D^-_s | (\overline{s} c)_L | 0 \rangle ,
\label{eq:Amp-DDs_f}
\end{eqnarray}

\begin{eqnarray}
{\cal A}_{\fac} (\overline{B}^0 \rightarrow D^+ D_s^{*-}) & = &
\frac{G_F}{\sqrt{2}} \left[ \rule{0mm}{6mm} V_{cb} V_{cs}^* \,
a_1 + \sum_{q = u, c} V_{qb} V_{qs}^* \left(a_4^q + 2 a^q_6
\frac{m_{D^*_s}^2} {(m_b - m_c) (m_c - m_s)} \right) \right] \nonumber\\
&& \times \; \langle D^+ | (\overline{c} b)_L |
\overline{B}^0 \rangle \, \langle D_s^{*-} | (\overline{s} c)_L | 0 \rangle,
\label{eq:Amp-DDsstar_f}
\end{eqnarray}

\begin{eqnarray}
{\cal A}_{\fac} (\overline{B}^0 \rightarrow D^{*+} D_s^-) & = &
\frac{G_F}{\sqrt{2}} \left[ \rule{0mm}{6mm} V_{cb} V_{cs}^* \, a_1 + \sum_{q = u, c} V_{qb} V_{qs}^* \left(a_4^q - 2 a^q_6
\frac{m_{D_s}^2}{(m_b + m_c) (m_c + m_s)} \right) \right] \nonumber\\
&& \times \; \langle D^{*+} | (\overline{c} b)_L |
\overline{B}^0 \rangle \, \langle D_s^- | (\overline{s} c)_L | 0 \rangle
\label{eq:Amp-DstarDs_f}
\end{eqnarray}
and

\begin{eqnarray}
{\cal A}_{\fac} (\overline{B}^0 \rightarrow D^{*+} D_s^{*-}) & = &
\frac{G_F}{\sqrt{2}} \left[ \rule{0mm}{6mm} V_{cb} V_{cs}^* \, a_1 
+ \sum_{q = u, c} V_{qb} V_{qs}^* \left(a_4^q + 2 a^q_6
\frac{m_{D^*_s}^2}{(m_b - m_c) (m_c - m_s)} \right) \right] \nonumber\\
&& \times \; \langle D^{*+} | \overline{c}
\gamma_\mu b | \overline{B}^0 \rangle \, \langle D_s^- | (\overline{s}
c)_L | 0 \rangle \nonumber\\
&& - \; \frac{G_F}{\sqrt{2}} \left[ \rule{0mm}{6mm} V_{cb} V_{cs}^* \, a_1
+ \sum_{q = u, c} V_{qb} V_{qs}^* \left(a_4^q -
2 a^q_6 \frac{m_{D^*_s}^2}{(m_b + m_c) (m_c - m_s)} \right) \right] \nonumber\\
&& \times \; \langle D^{*+} | \overline{c}
\gamma_\mu \gamma_5 b | \overline{B}^0 \rangle\, \langle D_s^- |
(\overline{s} c)_L | 0 \rangle ,
\label{eq:Amp-Dstar-Dsstar_f}
\end{eqnarray}
where
\begin{eqnarray}
a^q_3 & = & C^q_3 + {1 \over 3} C^q_4 , \nonumber\\
a^q_4 & = & C^q_4 + {1 \over 3} C^q_3 , \nonumber\\
a^q_5 & = & C^q_5 + {1 \over 3} C^q_6 , \nonumber\\
a^q_6 & = & C^q_6 + {1 \over 3} C^q_5 , \quad q = u, c .
\end{eqnarray}
Note that $N_c = 3$ is used for the penguin amplitudes also.

If the penguin-generated terms were omitted, the effect of working in NLL instead of LL would be very small. In fact, $\Delta {\cal B}_{\fac}$ is less than $1\%$, the same as that calculated above for the color-favored decays of type $b \rightarrow c \bar{u} d$.  However, if the contributions from the penguin diagrams are considered we get relatively large effects. The branching ratios for the processes $B \rightarrow D^+ D_s^-$ and $B \rightarrow D^+ D_s^{*-}$ get reduced by $27\%$ and $36\%$, respectively, while the branching ratios for $B \rightarrow D^* D_s$ and $B \rightarrow D^* D_s^*$ are increased by $6\%$ and $15\%$, respectively (see Table~\ref{tab:NLL-Ping-fac}). To demonstrate the cause of these large changes, let us consider the decay $B \rightarrow D^+ D_s^-$. In a rough calculation, we substitute the following approximations in (\ref{eq:Amp-DDs_f}): $a_1 \approx 1$, $a^c_4 \approx -0.04$, $a^c_6 \approx -0.05$, $V_{ub} V_{us}^* \approx 0$ and $m_{D_s}^2/(m_b - m_c) (m_c + m_s) \approx 1$. The change in the amplitude due to penguins is then about $(1 - 0.04 - 2 \times 0.05)^2 - 1 \approx -27\%$.

The decay amplitudes of class~II processes are simpler to derive. For example, the factorized decay amplitude for the process $\overline{B}^0 \rightarrow \overline{K}^0 J/\psi$ is given by
\begin{equation}
{\cal A}_{\fac} (\overline{B}^0 \rightarrow \overline{K}^0 J/\psi) = \frac{G_F}{\sqrt{2}} \left[ V_{cb} V_{cs}^* \, a_2 + \sum_{q = u, c} V_{qb} V_{qs}^* \left(a_3^q + a^q_5 \right) \right] \; \langle \overline{K}^0 | (\overline{s} b)_L | \overline{B}^0 \rangle \, \langle J/\psi | (\overline{c} c)_L | 0 \rangle .
\label{eq:Amp-K-PSI_f}
\end{equation}
Other processes of this class have similar amplitude forms. When the Wilson coefficients generated by the penguin diagrams are set to zero, the branching ratios calculated in NLL are $57\%$ lower than those calculated in LL (similar to class~II processes of type $b \rightarrow c \bar{u} d$). The penguin effects, however, turn out to be very small (about $0.2\%$). This is because the values for $a_3$ and $a_5$ are very close in magnitude and have opposite signs resulting in a mutual cancellation.


\subsection{Type $b \rightarrow c \bar{c} d$ Decays}

The effective Hamiltonian for processes of type $b \rightarrow c \bar{c} d$ is similar to Eq.~(\ref{eq:Heff-bccs-QCD}) except that the $s$ flavor is replaced by the $d$ flavor. So, the effective Wilson coefficients are calculated from Eqs.~(\ref{eq:C1-C6-NLL-eff}), using the following expression for $C_p$
\begin{eqnarray}
C_p & = & C_3(\mu) \left[ {4 \over 3} + {2 \over 3} \ln \frac{m_d^2}{\mu^2}  + {2 \over 3} \ln \frac{m_b^2}{\mu^2}  - \Delta F_1 \left(\frac{k^2}{m_d^2}\right) - \Delta F_1 \left(\frac{k^2}{m_b^2}\right)\right]
[C_4(\mu) + C_6(\mu)] \nonumber\\
&& \times \sum_{i = u, d, s, c, b} \left[ {2 \over 3} \ln \frac{m_i^2}{\mu^2} - \Delta F_1 \left(\frac{k^2}{m_i^2}\right) \right] .
\end{eqnarray}
The numerical values for $C_{i = 1, \ldots, 6}^{\mbox{\scriptsize eff}}$ turn out to be the same as those in Table~\ref{tab:Ceff-NLL} without noticeable changes. The decay amplitudes for the processes $B \rightarrow D D^-$, $D D^{*-}$, $D^* D^-$ and $D^* D^{*-}$ can be written from Eqs.~(\ref{eq:Amp-DDs_f})-(\ref{eq:Amp-Dstar-Dsstar_f}) by replacing the $s$~flavor by the $d$~flavor. By doing a similar replacement, the amplitude for the process $B \rightarrow \pi J/\psi$ can be written from Eq.~(\ref{eq:Amp-K-PSI_f}). In Table~\ref{tab:NLL-Ping-fac}, we show the NLL and Penguin effects on the calculated branching ratios of these two sets of processes.


\section{Estimation of Nonfactorization Parameters}

In Ref.~\cite{ref:Neubert-97}, nonfactorization was parametrized through $\varepsilon_1$ and $\varepsilon_8$. These two parameters represent the size of the color-singlet and color-octet nonfactorizable diagrams relative to the factorizable one (see Fig.~\ref{fig:diagrams-I}). By assuming universality (process-independence) of these two parameters, we estimated their values in Ref.~\cite{ref:Shamali-99} for Cabibbo-favored $B$ decays. The estimate was done using the available experimental branching ratios for two sets of class~I processes ($\bar{B}^0 \rightarrow D^+ \pi^-, D^+ \rho^-, D^+ a_1^-, D^{*+} \pi^-, D^{*+} \rho^-, D^{*+} a_1^-$ and $B \rightarrow D D_s, D D_s^*, D^* D_s, D^* D_s^*$) and one set of class~II processes ($B \rightarrow K J/\psi, K \psi(2S), J/\psi, K^* \psi(2S)$). For the Wilson coefficients we \cite{ref:Shamali-99} used the values calculated up to LL and neglected all contributions from the penguin diagrams. Regarding the form factors, we used the predictions of BSW~II model. In general, the estimated values of $\varepsilon_1$ and $\varepsilon_8$ improved the agreement with experimental measurements when the notion of factorization was extended to include other channels of $B$ and $B_s$~decays. This supported the assumption of universality of these parameters in Cabibbo-favored $B$ decays.

According to Eq.~(\ref{eq:a1eff-a2eff}), nonfactorization contribute to the decay amplitudes of class~I and class~II processes through the multiplicative facors (omitting the penguin contributions for simplicity of argument)
\begin{equation}
\xi_1 = \left( 1 + \varepsilon_1 + \frac{C_2}{a_1} \,\varepsilon_8 \right)
\approx \left( 1 + \varepsilon_1 - \frac{1}{3} \,\varepsilon_8 \right) \label{eq:a1eff}
\end{equation}
and 
\begin{equation}
\xi_2 = \left( 1 + \varepsilon_1 + \frac{C_1}{a_2} \,\varepsilon_8 \right)
\approx \left( 1 + \varepsilon_1 + 20 \,\varepsilon_8 \rule{0mm}{5mm}\right),
\label{eq:a2eff}
\end{equation}
respectively. The difference between these two factors is in the coefficient of the color-octet parameter (see Table~\ref{tab:Ceff-NLL}). Consequently, in case of class~II processes the long distance effects in $\varepsilon_8$ are greatly enhanced by the short distance effects arising from $C_1/a_2$. This is in addition to the enhancement of $\varepsilon_8$ over $\varepsilon_1$, by a factor of $N_c$, according to the rules of QCD. As a result, it may be harmless to ignore the contribution due to the color-singlet parameter $\varepsilon_1$. On the other hand, in class~I processes we notice that $\varepsilon_8$ is suppressed by approximately a factor of $1/3$ due to short-distance effects. Since this compensates for the enhancement due to QCD, it is not justified to omit $\varepsilon_1$ in this class and the two parameters should be treated on equal footing. From the preceding discussion we infer that class~II processes are sensitive probes of $\varepsilon_8$, whereas $\varepsilon_1$ is mainly determined by class~I processes, albeit with less sensitivity.

In this work, we re-estimate the values of the nonfactorization parameters $\varepsilon_1$ and $\varepsilon_8$ using a $\chi^2$ fit to the experimental branching ratios of the three sets of processes mentioned above. However, for the Wilson coefficients we use the values calculated up to NLL and include, in the color-singlet part of the amplitude, the contributions from the penguin diagrams. As for the color-octet part, we include the contributions from the operators $Q_1$ and $Q_2$ only. For example, the decay amplitude for the process $B \rightarrow K J/\psi$, after including the nonfactorizable contributions will read
\begin{equation}
{\cal A} (B \rightarrow K J/\psi) = \frac{G_F}{\sqrt{2}} V_{cb} V_{cs}^* \left[ \left(a_2 + \sum_{q = u, c} \frac{V_{qb} V_{qs}^*}{V_{cb} V_{cs}^*} \left(a_3^q + a^q_5 \right) \right) (1 +  \varepsilon_1) + C_1 \, \varepsilon_8 \right] \langle K | (\overline{s} b)_L | B \rangle \, \langle J/\psi | (\overline{c} c)_L | 0 \rangle .
\label{eq:ch6:Amp-K-PSI_nf}
\end{equation}

Since the nonfactorization parameters are model dependent, we repeat the $\chi^2$ fit using five different models for the form factors. The first of these models is the original Bauer, Stech and Wirbel model \cite{ref:Wirble-85} (called BSW~I here) where the form factors are calculated at zero momentum transfer and extrapolated using a monopole form for all the form factors. The second model (called BSW~II here) differs from the first one by using a dipole form to extrapolate the form factors $F_1$, $A_0$, $A_2$ and $V$. This is motivated by the consistency relations in the infinite quark mass limit derived in Ref.~\cite{ref:Neubert-92-2}. In the third model \cite{ref:Neubert-92-2} (called NRSX here) we use the heavy-quark effective theory predictions for the heavy-to-heavy form factors. For the heavy-to-light form factors we use the same values as those predicted by BSW~II model. In the fourth model (AW), developed by Altomari and Wolfenstein \cite{ref:Altomari-88}, the form factors are evaluated at the zero-recoil point corresponding to the maximum momentum transfer and then extrapolated down to the required momentum using a monopole form. In the last model (ISGW), by Isgur, Scora, Grinstein and Wise \cite{ref:Isgur-89}, the form factors are calculated at the maximum momentum transfer and extrapolated down with an exponential form. See Table~\ref{tab:form-factors} for a comparison between the form factors predicted by these models for the process $B \rightarrow D D_s$.

\begin{figure}
\epsfxsize=6.0in  \centerline{\epsffile{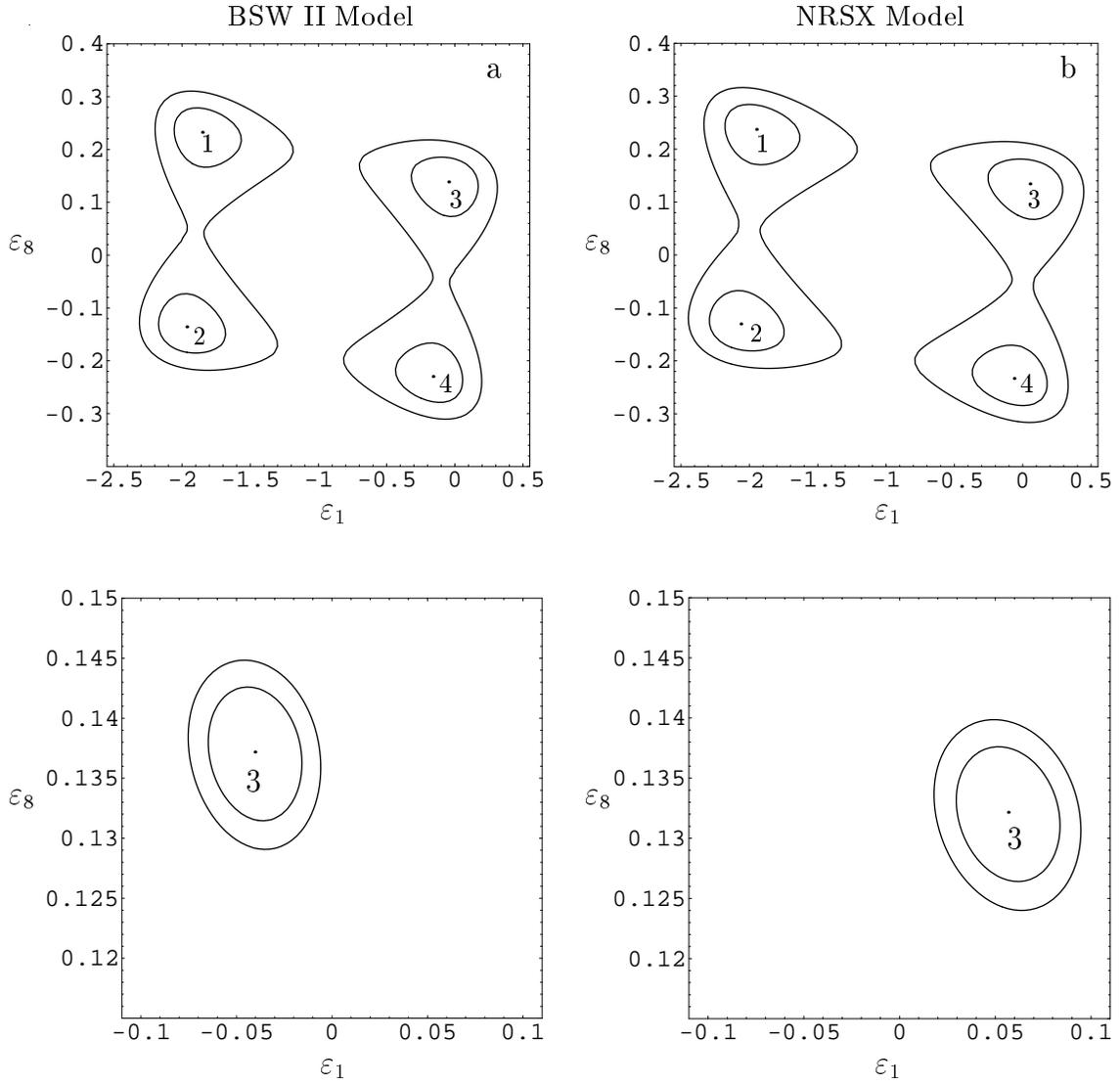}}

\caption{(a) A contour plot of $\chi^2$ in $\varepsilon_1$-$\varepsilon_8$ space using BSW~II model. (b) A contour plot of $\chi^2$ in $\varepsilon_1$-$\varepsilon_8$ space using NRSX model. The lower graphs show a magnification of the region containing minimum 3 of $\chi^2$ in the corresponding graphs. The inner closed curve represents a change of $\Delta \chi^2 = 1$ from the minimum while the outer closed curve  represents a change of $\Delta \chi^2 = 2$.}
\label{fig:ch6:x1-x8-NLL}
\end{figure}

Working, for example, in BSW~II model we show in Fig.~\ref{fig:ch6:x1-x8-NLL}~(a) a contour plot of $\chi^2$ in $\varepsilon_1$-$\varepsilon_8$ space. The four minima, labeled 1, 2, 3 and 4, that appear in this figure is to be compared with the corresponding regions in Fig.~2 of  Ref.\ \cite{ref:Shamali-99}. The value of $\chi^2$ per degree of freedom ($\chi^2/d$) for these minima is 0.6\@. In an argument similar to that used in Ref.~\cite{ref:Shamali-99}, we exclude solutions~1, and 2 due to the severe violation of the approximate relation $\varepsilon_1/\varepsilon_8 = \pm 1/N_c$ suggested by $1/N_c$ expansion. Also, solution~4 is excluded because it produces a negative value for the effective $a_2$ parameter. So, we end up with the estimate $\varepsilon_1 = - 0.040 \pm 0.024$ and $\varepsilon_8 = 0.137 \pm 0.006$. The uncertainties correspond to $\Delta \chi^2 = 1$. The predictions of the other models are shown in Table~\ref{tab:eps1-eps8-chi}.

Out of the five models considered we note that three (BSW~I, ISGW and AW) do not produce a good fit to the experimental data. This is indicated by the high values of $\chi^2/d$. The remaining two models, on the other hand, produce a good fit to the data with $\chi^2/d$ equal to 0.6 for BSW~II model and 0.4 for NRSX model. In these two models which show interesting fits the heavy-to-light form factors are calculated in the same way. So, the set of class~II, processes ($B \rightarrow K J/\psi$ \ldots etc.) have the same values for the form factors. This is reflected in close predictions of $\varepsilon_8$ by both models, which is not the case for the other parameter where the two sets of class~I processes take different values for the form factors in the two models. Another point to be noticed is that NRSX model predicts a destructive interference between the color-singlet and color-octet nonfactorization contributions causing almost a complete cancellation between the two for class~I processes. This is not the case for the other models which suggest a constructive interference.


\section{Predicted Decay Rates Including Nonfactorizable Contributions}

Using the values estimated for the nonfactorization parameters in a scheme with NLL Wilson coefficients and penguin contributions, we calculate the branching ratios, in BSW~II model, for all processes considered in the previous work \cite{ref:Shamali-99}. The results are shown in Tables~\ref{tab:B-decays-CF} and \ref{tab:Bs-decays-CF}. Also, by assuming that the universality of $\varepsilon_1$ and $\varepsilon_8$ extends to Cabibbo-suppressed processes, we evaluate the branching ratios for a set of class~I processes of type $b \rightarrow c \bar{c} d$ and for another set of class~II processes of the same type. The results are shown in Table~\ref{tab:B-decays-CS}. These two sets were not considered in the previous work. 

Our predictions show a good agreement with available experimental data which includes the recently measured decay channels $\overline{B}^0 \rightarrow D^{*+} D^{*-}$ \cite{ref:CLEO-99} and $B^- \rightarrow \pi^- J/\psi$ \cite{ref:PDG-98}. In the calculations, the two states $| \eta \rangle$ and $| \eta' \rangle$ are treated in the same way as in \cite{ref:Shamali-98} where the mixing angle and wavefunction normalizations are properly taken care of. As for the decay constants we used the values adopted in Ref.~\cite{ref:Shamali-99}.


\section{Discussion and Conclusion}

In Ref.~\cite{ref:Shamali-99}, we demonstrated that naive factorization, in LL and using tree diagrams only, gives reasonable predictions in comparison with experimental measurements for the branching ratios of class~I and class~III processes. However, for class~II processes the predicted branching ratios were very low. Also, it was demonstrated that including nonfactorizable contributions through the parameters $\varepsilon_1$ and $\varepsilon_8$ improves considerably the predicted branching ratios for the latter class while preserving the reasonable predictions for the other two.

\begin{figure}
\epsfxsize=3.4in  \centerline{\epsffile{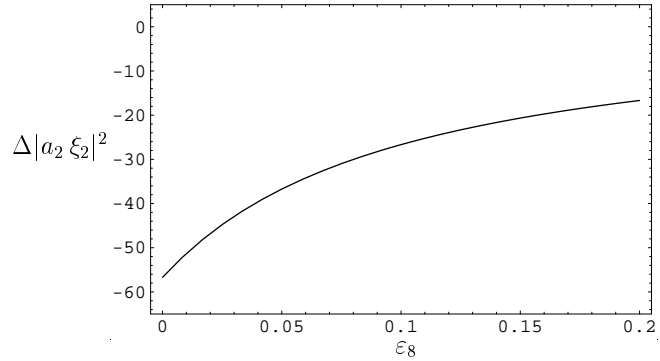}}

\caption{The percentage difference between $|a_2 \,\xi_2|^2_{LL}$ and $|a_2 \,\xi_2|^2_{NLL}$ as a function of $\varepsilon_8$, taking $\varepsilon_1 = 0$. }
\label{fig:delta-a2eff}
\end{figure}

In this work, we find that by working in NLL (with no penguins) the predicted branching ratios (in naive factorization) of class~I and class~III processes are very close to the LL predictions and to the experimental values (see Table~\ref{tab:NLL-Ping-fac}). For class~II processes, on the other hand, the predicted branching ratios are considerably lower (by about 57\%) than the LL predictions making the disagreement with the experimental values even worse. However, this problem is greatly remedied by including nonfactorizable contributions. Beside the enhancement of the branching ratios of class~II processes, the inclusion of nonfactorizable contributions reduces the sensitivity to whether LL or NLL Wilson coefficients are used in the calculation. This is demonstrated in Fig.~\ref{fig:delta-a2eff} by plottting the percentage difference between $|a_2 \,\xi_2|^2_{LL}$ and $|a_2 \,\xi_2|^2_{NLL}$ as a function of $\varepsilon_8$, taking $\varepsilon_1 = 0$. From the graph we see that a 10\% contribution to $\varepsilon_8$ reduces the difference between the branching ratios predicted by LL and NLL to about half that in naive factorization.

Penguin diagrams contribute to processes of types $b \rightarrow c \bar{c} s$
and $b \rightarrow c \bar{c} d$. In both types, class~II processes are affected only slightly by the penguin contributions. This is due to the destructive interference between the different terms, in the amplitude, generated by the penguins. On the other hand, for class~I processes of type $b \rightarrow c \bar{c} s$ this cancellation does not happen and the decay amplitudes receive significant contribution from the penguin diagrams (see Table~\ref{tab:NLL-Ping-fac}).

As can be seen from Table~\ref{tab:eps1-eps8-chi} the best fit to the experimental data is produced by the Heavy-Quark Effective Theory (contained in the NRSX model), and lends support to the assumption of universality of the nonfactorizable contributions in $B$~decays.


\section*{Acknowledgement}

This research was supported by a grant to A.N.K. from the Natural Sciences and  Engineering Research Council of Canada.



\begin{table}
\caption{The Wilson Coefficients in LL and NLL using the NDR. All values are evaluated at the scale $\mu = 4.6 \,\mbox{GeV}$, and for $\Lambda^5_{\MS}= 219\,\mbox{MeV}$.}
\label{tab:Wilson-LL-NLL}

\begin{tabular}{lrr}
& LL & NLL \\ \hline
$C_1(\mu)$ & $1.127$ & $1.075$ \\
$C_2(\mu)$ & $-0.286$ & $-0.178$ \\
$C_3(\mu)$ & $0.013$ & $0.012$ \\
$C_4(\mu)$ & $-0.029$ & $-0.033$ \\
$C_5(\mu)$ & $0.008$ & $0.009$ \\
$C_6(\mu)$ & $-0.037$ & $-0.039$ \\\hline
$a_1(\mu)$ & $1.032$ & $1.016$ \\
$a_2(\mu)$ & $0.090$ & $0.180$ \\
$C_2/a_1$ & $-0.277$ & $-0.175$ \\
$C_1/a_2$ & $12.54$ & $5.96$
\end{tabular}

\end{table}

\begin{table}
\caption{Effective Wilson Coefficients in NLL, evaluated using
$\Lambda^5_{\MS} = 219 \,\mbox{MeV}$ and the running quark masses at the $b$-quark mass scale.}
\label{tab:Ceff-NLL}

\begin{tabular}{lcc}
$C_1^{\eff}$ & $1.143$ &  \\
$C_2^{\eff}$ & $-0.322$ & \\ \hline
& $q = u$ &  $q = c$ \\
$C_3^{q\,\eff}$ & $0.0184 + 0.0048\,i$ & $0.0197 + 0.0044\,i$ \\
$C_4^{q\,\eff}$ & $-0.0407 - 0.0145\,i$ & $-0.0453 - 0.0132\,i$ \\
$C_5^{q\,\eff}$ & $0.0130 + 0.0045\,i$ & $0.0145 + 0.0044\,i$ \\
$C_6^{q\,\eff}$ & $-0.0522 - 0.0145\,i$ & $-0.0568 - 0.0132\,i$ \\\hline
$a_1$ & $1.036$ & \\
$a_2$ & $0.059$ & \\
$C_2/a_1$ & $-0.311$ & \\
$C_1/a_2$ & $19.37$ &
\end{tabular}

\end{table}

\begin{table}
\centering
\caption{$\Delta {\cal B}_{\fac}$ represents the percentage change in the branching ratio, assuming factorization, due to NLL values of the Wilson coefficients (column 4) and due to penguin diagrams (column 5). The last column represents the total change.}

\begin{tabular}{lccrrr}
Processes & Class & Model & \multicolumn{3}{c}{$\Delta {\cal B}_{\fac}$: Change in branching ratio} \\
&&& NLL effect & Penguin effect & Total change\\\hline
\multicolumn{6}{l}{\bf Type $b \rightarrow c \bar{u} d$} \\
$\overline{B}^0 \rightarrow D^+ \pi^-$ \ldots etc. & I && $0.8\%$ && $0.8\%$ \\
$\overline{B}^0 \rightarrow D^0 \pi^0$ \ldots etc. & II && $-56.7\%$ && $-56.7\%$ \\
$B^- \rightarrow D^0 \pi^-$ & III & BSW~II & $-4.4\%$ & & $-4.4\%$ \\
$B^- \rightarrow D^0 \rho^-$ & III & BSW~II & $-2.1\%$ & & $-2.1\%$ \\
$B^- \rightarrow D^0 a^-_1$ & III & BSW~II & $-1.1\%$ & & $-1.1\%$ \\
$B^- \rightarrow D^{*0} \pi^-$ & III & BSW~II & $-6.0\%$ & & $-6.0\%$ \\
$B^- \rightarrow D^{*0} \rho^-$ & III & BSW~II & $-3.3\%$ & & $-3.3\%$ \\
$B^- \rightarrow D^{*0} a^-_1$ & III & BSW~II & $-1.8\%$ & & $-1.8\%$ \\\hline
\multicolumn{6}{l}{\bf Type $b \rightarrow c \bar{c} s$} \\
$B \rightarrow D D_s$ & I && $0.8\%$ & $-27\%$ & $-26.2\%$ \\
$B \rightarrow D D_s^*$ & I && $0.8\%$ & $-36.0\%$ & $-35.2\%$ \\
$B \rightarrow D^* D_s$ & I && $0.8\%$ & $6.4\%$ & $7.2\%$ \\
$B \rightarrow D^* D_s^*$ & I & BSW~II & $0.8\%$ & $15.1\%$ & $15.9\%$ \\
$B \rightarrow K J/\psi$ \ldots etc. & II && $-56.7\%$ & $0.2\%$ & $-56.5\%$ \\\hline
\multicolumn{6}{l}{\bf Type $b \rightarrow c \bar{c} d$} \\
$B \rightarrow D D^-$ & I && $0.8\%$ & $-23\%$ & $-22.2\%$ \\
$B \rightarrow D D^{*-}$ & I && $0.8\%$ & $-25.7\%$ & $-24.9 \%$ \\
$B \rightarrow D^* D^-$ & I && $0.8\%$ & $5.9\%$ & $6.7\%$ \\
$B \rightarrow D^* D^{*-}$ & I & BSW~II & $0.8\%$ & $12.9\%$ & $13.7\%$ \\
$B \rightarrow \pi J/\psi$ \ldots etc. & II && $-56.7\%$ &  $0.4\%$ & $-56.3\%$
\end{tabular}

\label{tab:NLL-Ping-fac}
\end{table}

\begin{table}
\centering
\caption{Predictions of the $F_0$ and $F_1$ form factors for the process $B \rightarrow D D_s$ in the five models considered in this work.}

\begin{tabular}{lcc}
Model & $F_0(m^2_{D_s})$ & $F_1(m^2_{D_s})$ \\\hline
BSW~I \cite{ref:Wirble-85} & 0.756 & 0.767 \\
BSW~II \cite{ref:Wirble-85,ref:Neubert-92} & 0.756 & 0.849 \\
NRSX \cite{ref:Neubert-92-2} & 0.653 & 0.725 \\
AW \cite{ref:Altomari-88} & 0.889 & 0.937 \\
ISGW \cite{ref:Isgur-89} & 0.832 & 0.900
\end{tabular}

\label{tab:form-factors}
\end{table}

\begin{table}
\centering
\caption{The estimated nonfactorization parameters $\varepsilon_1$ and $\varepsilon_8$ in different models. The last column shows the value of $\chi^2$ per degree of freedom which indicate the goodness of fit.}

\begin{tabular}{lccc}
Model & $\varepsilon_1$ & $\varepsilon_8$ & $\chi^2/d$ \\\hline
BSW~I \cite{ref:Wirble-85} & $-0.037 \pm 0.024$ & $0.139 \pm 0.006$ & $5.8$ \\
BSW~II \cite{ref:Wirble-85,ref:Neubert-92} & $-0.040 \pm 0.024$ & $0.137 \pm 0.006$ & $0.6$ \\
NRSX \cite{ref:Neubert-92-2} & $0.057 \pm 0.027$ & $0.132 \pm 0.006$ & $0.4$ \\
AW \cite{ref:Altomari-88} & $-0.250 \pm 0.019$ & $0.171 \pm 0.007$ & $5.7$ \\
ISGW \cite{ref:Isgur-89} & $-0.071 \pm 0.024$ & $0.230 \pm 0.008$ & $2.3$
\end{tabular}

\label{tab:eps1-eps8-chi}
\end{table}

\begin{table}
\centering
\caption{The branching ratios predicted for a number of Cabibbo-favored $B$ decays in BSW~II model. The values in the second column were calculated by taking $\varepsilon_1(\mu_0) = \varepsilon_8(\mu_0) = 0$ whereas the values  in the third  column were calculated by taking $\varepsilon_1(\mu_0) = - 0.040 \pm 0.024$ and $\varepsilon_8(\mu_0) = 0.137 \pm 0.006$. The last column represents the available experimental measurements.}

\begin{tabular}{lccc} 
Process & Fac. & Nonfac. & Exp. \cite{ref:PDG-98} \\\hline
& \multicolumn{3}{c}{Branching Ratio $\times 10^{-3}$} \\
$\bar{B}^0 \rightarrow D^+ \pi^-$ & $4.1 \pm 0.1$ & $3.4 \pm 0.2$ & $3.0 \pm 0.4$ \\
$\bar{B}^0 \rightarrow D^+ \rho^-$ & $9.9 \pm 0.3$ & $8.4 \pm 0.5$ & $7.9 \pm 1.4$ \\
$\bar{B}^0 \rightarrow D^+ a_1^-$ & $11.3 \pm 1.0$ & $9.5 \pm 1.0$ & $6.0 \pm 3.3$ \\
$\bar{B}^0 \rightarrow D^{*+} \pi^-$ & $3.2 \pm 0.1$ & $2.6 \pm 0.2$ &  $2.76 \pm 0.21$\\
$\bar{B}^0 \rightarrow D^{*+} \rho^-$ & $9.0 \pm 0.2$ & $7.6 \pm 0.4$ & $6.7 \pm 3.3$\\
$\bar{B}^0 \rightarrow D^{*+} a_1^-$ & $12.9 \pm 1.2$ & $10.9 \pm 1.2$  &  $13.0 \pm 2.7$\\ \hline
& \multicolumn{3}{c}{Branching Ratio $\times 10^{-4}$} \\
$\bar{B}^0 \rightarrow D^0 \pi^0$ & $0.06 \pm 0.01$ & $0.77 \pm 0.16$ & $< 1.2$ \\
$\bar{B}^0 \rightarrow D^{*0} \pi^0$ & $0.08 \pm 0.02$ & $1.09 \pm 0.23$ & $< 4.4$ \\
$\bar{B}^0 \rightarrow D^0 \eta$ & $0.03 \pm 0.01$ & $0.43 \pm 0.09$ & $< 1.3$ \\
$\bar{B}^0 \rightarrow D^{*0} \eta$ & $0.05 \pm 0.01$ & $0.59 \pm 0.13$ & $< 2.6$ \\
$\bar{B}^0 \rightarrow D^0 \eta'$ & $0.01 \pm 0.01$ & $0.13 \pm 0.03$ & $< 9.4$ \\
$\bar{B}^0 \rightarrow D^{*0} \eta'$ & $0.01 \pm 0.01$ & $0.18 \pm 0.04$ & $< 14$ \\
$\bar{B}^0 \rightarrow D^0 \rho^0$ & $0.04 \pm 0.01$ & $0.53 \pm 0.11$ & $< 3.9$ \\
$\bar{B}^0 \rightarrow D^{*0} \rho^0$ & $0.09 \pm 0.02$ & $1.13 \pm 0.24$ & $< 5.6$ \\
$\bar{B}^0 \rightarrow D^0 \omega$ & $0.04 \pm 0.01$ & $0.52 \pm 0.11$ & $< 5.1$ \\
$\bar{B}^0 \rightarrow D^{*0} \omega$ & $0.09 \pm 0.02$ & $1.12 \pm 0.24$ & $< 7.4$ \\ \hline
& \multicolumn{3}{c}{Branching Ratio $\times 10^{-3}$} \\
$B^- \rightarrow D^0 \pi^-$ & $4.8 \pm 0.1$ & $5.4 \pm 0.3$ & $5.3 \pm 0.5$ \\
$B^- \rightarrow D^0 \rho^-$ & $11.1 \pm 0.3$ & $10.9 \pm 0.7$ & $13.4 \pm 1.8$ \\
$B^- \rightarrow D^0 a_1^-$ & $12.4 \pm 1.1$ & $11.6 \pm 1.2$ & \\
$B^- \rightarrow D^{*0} \pi^-$ & $3.8 \pm 0.1$ & $4.6 \pm 0.3$ & $4.6 \pm 0.4$ \\
$B^- \rightarrow D^{*0} \rho^-$ & $10.3 \pm 0.3$ & $10.9 \pm 0.3$ &  $15.5 \pm 3.1$\\
$B^- \rightarrow D^{*0} a_1^-$ & $14.4 \pm 0.4$ & $13.9 \pm 0.4$ & \\ \hline
& \multicolumn{3}{c}{Branching Ratio $\times 10^{-3}$} \\
$B \rightarrow D D_s$ & $10.3 \pm 2.2$ & $8.5 \pm 1.9$ & $9.8 \pm 2.4$ \\
$B \rightarrow D D_s^*$ & $7.9 \pm 1.6$ & $6.5 \pm 1.4$ & $9.4 \pm 3.1$\\
$B \rightarrow D^* D_s$ & $8.4 \pm 1.8$ & $7.1 \pm 1.6$ & $10.4 \pm 2.8$\\
$B \rightarrow D^* D_s^*$ & $29.4 \pm 5.9$ & $24.9 \pm 5.2$ & $22.3 \pm 5.7$ \\ \hline
& \multicolumn{3}{c}{Branching Ratio $\times 10^{-4}$} \\
$B \rightarrow K J/\psi$ & $0.72 \pm 0.05$ & $9.4 \pm 0.9$ & $9.5 \pm 0.8$ \\
$B \rightarrow K \psi(2S)$ & $0.38 \pm 0.04$ & $5.0 \pm 0.6$ & $5.8 \pm 1.2$ \\
$B \rightarrow K^* J/\psi$ & $1.17 \pm 0.09$ & $15.2 \pm 1.5$ & $14.6 \pm 1.4$ \\
$B \rightarrow K^* \psi(2S)$ & $0.70 \pm 0.07$ & $9.1 \pm 1.1$ & $9.6 \pm 2.5$
\end{tabular}

\label{tab:B-decays-CF}
\end{table}

\begin{table}
\centering
\caption{The branching ratios predicted for a number of Cabibbo-favored $B_s$ decays in BSW~II model. The values in the second column were calculated by taking $\varepsilon_1(\mu_0) = \varepsilon_8(\mu_0) = 0$ whereas the values  in the third  column were calculated by taking $\varepsilon_1(\mu_0) = - 0.040 \pm 0.024$ and $\varepsilon_8(\mu_0) = 0.137 \pm 0.006$. The last column represents the available experimental measurements.}

\begin{tabular}{lccc} 
Process & Fac. & Nonfac. & Exp. \cite{ref:PDG-98} \\\hline
& \multicolumn{3}{c}{Branching Ratio $\times 10^{-3}$} \\
$\bar{B}_s \rightarrow D_s^+ \pi^-$ & $3.6 \pm 0.2$ & $3.0 \pm 0.2$ & $<130$ \\
$\bar{B}_s \rightarrow D_s^+ \rho^-$ & $8.7 \pm 0.4$ & $7.3 \pm 0.5$ & \\
$\bar{B}_s \rightarrow D_s^+ a_1^-$ & $9.9 \pm 1.0$ & $8.3 \pm 0.9$ & \\
$\bar{B}_s \rightarrow D_s^{*+} \pi^-$ & $2.6 \pm 0.1$ & $2.2 \pm 0.2$ & \\
$\bar{B}_s \rightarrow D^{*+} \rho^-$ & $7.6 \pm 0.4$ & $6.4 \pm 0.5$ & \\
$\bar{B}_s \rightarrow D^{*+} a_1^-$ & $10.8 \pm 1.1$ & $9.1 \pm 1.0$  &  \\ \hline
& \multicolumn{3}{c}{Branching Ratio $\times 10^{-4}$} \\
$\bar{B}_s \rightarrow D^0 K^0$ & $0.08 \pm 0.02$ & $1.0 \pm 0.2$ & \\
$\bar{B}_s \rightarrow D^{*0} K^0$ & $0.11 \pm 0.02$ & $1.4 \pm 0.3$ & \\
$\overline{B}_s \rightarrow D^0 K^{*0}$ & $0.06 \pm 0.01$ & $0.7 \pm 0.2$ & \\
$\bar{B}_s \rightarrow D^{*0} K^{*0}$ & $0.12 \pm 0.02$ & $1.5 \pm 0.3$ & \\ \hline
& \multicolumn{3}{c}{Branching Ratio $\times 10^{-4}$} \\
$\bar{B}_s \rightarrow \eta J/\psi$ & $0.19 \pm 0.02$ & $2.5 \pm 0.3$ & $< 38$ \\
$\bar{B}_s \rightarrow \eta \psi(2S)$ & $0.10 \pm 0.01$ & $1.4 \pm 0.2$ & \\
$\bar{B}_s \rightarrow \eta' J/\psi$ & $0.22 \pm 0.02$ & $2.9 \pm 0.3$ & \\
$\bar{B}_s \rightarrow \eta' \psi(2S)$ & $0.10 \pm 0.01$ & $1.3 \pm 0.2$ & \\
$\bar{B}_s \rightarrow \phi J/\psi$ & $0.79 \pm 0.07$ & $10.3 \pm 1.1$ & $9.3 \pm 3.3$ \\
$\bar{B}_s \rightarrow \phi \psi(2S)$ & $0.47 \pm 0.05$ & $6.2 \pm 0.8$ &
\end{tabular}

\label{tab:Bs-decays-CF}
\end{table}

\begin{table}
\centering
\caption{The branching ratios predicted for a number of Cabibbo-suppressed $B$ decays in BSW~II model. The values in the second column were calculated by taking $\varepsilon_1(\mu_0) = \varepsilon_8(\mu_0) = 0$ whereas the values  in the third  column were calculated by taking $\varepsilon_1(\mu_0) = - 0.040 \pm 0.024$ and $\varepsilon_8(\mu_0) = 0.137 \pm 0.006$. The last column represents the available experimental measurements.}

\begin{tabular}{lccc}
Process & Fac. & Nonfac. & Exp. \\\hline
& \multicolumn{3}{c}{Branching Ratio $\times 10^{-4}$} \\
$\overline{B}^0 \rightarrow D^+ D^-$ & $3.6 \pm 0.7$ & $3.0 \pm 0.6$ & \\
$\overline{B}^0 \rightarrow D^+ D^{*-}$ & $3.5 \pm 0.7$ & $2.9 \pm 0.6$ & \\
$\overline{B}^0 \rightarrow D^{*+} D^-$ & $2.7 \pm 0.6$ & $2.4 \pm 0.5$ & \\
$\overline{B}^0 \rightarrow D^{*+} D^{*-}$ & $9.9 \pm 2.0$ & $8.3 \pm 1.7$ & $6.2 \pm 3.6$ \cite{ref:CLEO-99} \\ \hline
& \multicolumn{3}{c}{Branching Ratio $\times 10^{-5}$} \\
$B^- \rightarrow \pi^- J/\psi$ & $0.35 \pm 0.03$ & $4.6 \pm 0.5$ & $5.0 \pm 1.5$ \cite{ref:PDG-98} \\
$B^- \rightarrow \rho^- J/\psi$ & $0.51 \pm 0.04$ & $6.7 \pm 0.7$ & $< 77$ \\
$B^- \rightarrow a_1^- J/\psi$ & $0.23 \pm 0.02$ & $3.0 \pm 0.3$ & $< 120$ \\
$B^- \rightarrow \pi^- \psi(2S)$ & $0.21 \pm 0.02$ & $2.8 \pm 0.3$ & \\
$B^- \rightarrow \rho^- \psi(2S)$ & $0.33 \pm 0.03$ & $4.4 \pm 0.5$ & \\
$B^- \rightarrow a_1^- \psi(2S)$ & $0.11 \pm 0.01$ & $1.5 \pm 0.2$ &
\end{tabular}

\label{tab:B-decays-CS}
\end{table}

\end{document}